\documentclass[12pt]{article}
\usepackage{graphicx}
\usepackage{amssymb}

\newcommand{\be}{\begin{equation}}
\newcommand{\ee}{\end{equation}}
\begin{document}
\topmargin 0pt
\oddsidemargin=-0.4truecm
\evensidemargin=-0.4truecm
\renewcommand{\thefootnote}{\fnsymbol{footnote}}
\newpage
\setcounter{page}{0}
\begin{titlepage}
\vspace*{-2.0cm}
\begin{flushright}
\end{flushright}
\vspace*{0.1cm}
\begin{center}
{\Large \bf Improving LMA predictions with non standard interactions
} \\
\vspace{0.6cm}
\vspace{0.4cm}

{\large
C. R. Das\footnote{E-mail: crdas@cftp.ist.utl.pt},
Jo\~{a}o Pulido\footnote{E-mail: pulido@cftp.ist.utl.pt}\\
\vspace{0.15cm}
{  {\small \sl CENTRO DE F\'{I}SICA TE\'{O}RICA DAS PART\'{I}CULAS (CFTP) \\
 Departamento de F\'\i sica, Instituto Superior T\'ecnico \\
Av. Rovisco Pais, P-1049-001 Lisboa, Portugal}\\
}}
\vspace{0.25cm}
\end{center}
\vglue 0.6truecm

\begin{abstract}
It has been known for some time that the well established LMA solution to the observed 
solar neutrino deficit fails to predict a flat energy spectrum for SuperKamiokande as 
opposed to what the data indicates. It also leads to a Chlorine rate which appears to 
be too high as compared to the data. We investigate the possible solution to these
inconsistencies with non standard neutrino interactions, assuming that they come as 
extra contributions to the $\nu_{\alpha}\nu_{\beta}$ and $\nu_{\alpha}e$ vertices that 
affect both the propagation of neutrinos through solar matter and their detection. We find
that, among the many possibilities for non standard couplings, only one of them leads to a 
flat SuperKamiokande spectral rate in better agreement with the data and predicts a Chlorine
rate within 1$\sigma$ of the observed one, while keeping all other predictions accurate. 
\end{abstract}

\end{titlepage}

\section{Introduction}

Neutrino non-standard interactions (NSI) have been introduced long ago \cite{Guzzo:1991hi,
Roulet:1991sm} to account for a possible alternative solution to the solar neutrino 
problem. Since then a great deal of effort has been dedicated to study its possible
consequences. To this end possible NSI signatures in neutrino processes have been investigated, 
models for neutrino NSI have been developed and bounds have been derived \cite{Grossman:1995wx,
Johnson:1999ci,Datta:2000ci,Huber:2001zw,Huber:2001de,Ota:2001pw,Huber:2002bi,Davidson:2003ha,
Barranco:2005ps,Mangano:2006ar,Antusch:2008tz,Biggio:2009kv,Gago:2009ij,Biggio:2009nt,Wei:2010ww}.
Specific investigations of NSI in matter have also been performed within the context
of supernova \cite{EstebanPretel:2007yu} and solar neutrinos \cite{Berezhiani:2001rt,
Friedland:2004pp,Guzzo:2004ue,Miranda:2004nb,Bolanos:2008km,Escrihuela:2009up}.

Although LMA is generally accepted as the dominant solution to the solar neutrino problem 
\cite{Fogli:2008ig,Schwetz:2008er}, not only its robustness has been challenged by NSI, as it 
can shift the LMA solution to the dark side region of parameter space \cite{Miranda:2004nb},
but also some inconsistencies remain regarding its agreement with the data \cite{Das:2009kw,
Pulido:2009sb}. In fact, while the SuperKamiokande (SK) energy spectrum appears to be flat
\cite{Fukuda:2002pe,:2008zn}, the LMA prediction shows a clear negative slope in the same energy 
range. With the expected improvement in the trigger efficiency for threshold electron energies 
as low as $3~MeV$ to be reached in the near future \cite{Smy}, such a disagreement, if it 
persists, may become critical. Moreover the LMA solution predicts an event rate for the 
Cl experiment \cite{Cleveland:1998nv} which is 2$\sigma$ above the observed one 
\cite{deHolanda:2003tx}. These are motivations to consider 'beyond LMA' solar neutrino 
solutions in which NSI may play a subdominant, although important role.

In order for NSI to be detectable and therefore relevant in physical processes, the 
characteristic scale of the new physics must not be too much higher than the scale of
the physics giving rise to the Standard Model interactions, $\Lambda_{EW}\simeq G_F^{-1/2}$. 
Possible realisations are one loop radiative models of Majorana neutrino mass 
\cite{AristizabalSierra:2007nf}, supersymmetric SO(10) with broken D-parity 
\cite{Malinsky:2005bi}, the inverse seesaw in a supersymmetry context 
\cite{Bazzocchi:2009kc} or triplet seesaw models \cite{Malinsky:2008qn}. 
Since the scale at which the new interaction arises is supposed to be not too far 
from the electroweak scale, its coupling may be parameterised by $G_F\varepsilon$ 
where $\varepsilon\simeq \Lambda^2_{EW}/\Lambda^2_{NP}\simeq 10^{-2}$ for d=6 or 
$\varepsilon\simeq \Lambda^4_{EW}/\Lambda^4_{NP}\simeq 10^{-4}$ for d=8 operators 
respectively. For type I seesaw, NSI are of course negligible.

In this paper we will be concerned with NSI both at the level of propagation
through solar matter and at the level of detection. Matter NSI are defined through 
the addition of an effective operator to the Lagrangian density   
\begin{equation}
{\cal L}^{M}_{NSI}=-2\sqrt{2}G_F \varepsilon^{fP}_{\alpha \beta}[\bar f\gamma^{\mu}
P f][\bar\nu_{\alpha}\gamma_{\mu}P_L \nu_{\beta}]
\end{equation}
where $f=e,u,d$, $P$ denotes the projection operator for left and right chirality 
and $\varepsilon^{fP}_{\alpha \beta}$ parameterizes the deviation from 
the standard interactions. At present there is no evidence at all of such operators
generated at a scale $\Lambda_{NP}$, hence the variety of theoretical models for the 
physics accessible to the LHC. 

There are several ways to introduce NSI. For instance in fermionic seesaw models, once 
the heavy fermions (singlets or triplets) are integrated out, modified couplings of 
leptons to gauge bosons are obtained in the form of a non-unitary leptonic matrix.
The strong bounds on the deviation from the unitarity of this matrix constrain these
NSI to be $\lesssim O(10^{-3})$ \cite{Antusch:2008tz}. Alternatively NSI can be generated 
by other new physics above the electroweak scale not related to neutrino masses. As a 
consequence an SU(2) gauge invariant formulation of NSI is required, since any gauge 
theory beyond the standard model must necessarily respect its gauge symmetry. 
Strong bounds from four charged fermionic processes \cite{Berezhiani:2001rs} and 
electroweak precision tests requiring fine tunings imply that possibilities are 
limited for such scenarios \cite{Biggio:2009kv,Gavela:2008ra}. Another way to introduce
NSI is by assuming that these are extra contributions to the vertices $\nu_{\alpha}
\nu_{\beta}$ and $\nu_{\alpha}e$. In such a case the parameters $\varepsilon_{\alpha
\beta}$ describe the deviation from the standard model vertices and are treated like
the standard interactions. It is possible that other effects are present in this case
that depend on the nature and number of particles that may be introduced in a particular
model. We adopt this procedure in the present paper and assume these model dependent 
effects to be negligible.

The paper is organized as follows: section 2 is devoted to the study of the propagation 
and detection of solar neutrinos. We start by reviewing the derivation of the neutrino 
refraction indices with standard interactions (SI) and their generalization to NSI in 
order to obtain the matter 
Hamiltonian. The survival and conversion probabilities to $\nu_{\mu}$ and $\nu_{\tau}$
are then evaluated through the numerical integration of the Schr\"{o}dinger like equation
using the Runge-Kutta method and the experimental event rates are obtained. We use the
reference solar model with high metalicity, BPS08(GS) \cite{PenaGaray:2008qe}. In section 3 
we investigate the influence of the NSI couplings on these rates in order to find whether
and how the fits can be improved with respect to the LMA ones. We concentrate in 
particular on the elimination of the upturn in the SuperKamiokande spectrum predicted by 
LMA for energies below 8-10 $MeV$ not supported by the data \cite{Das:2009kw,Pulido:2009sb,
Fukuda:2002pe,:2008zn} and on the Chlorine rate whose LMA prediction 
exceeds the data by 2$\sigma$ \cite{Cleveland:1998nv,deHolanda:2003tx}. 
We find that there are quite limited possibilities for the NSI couplings allowing for 
such fit improvements. Finally in section 4 we draw our main conclusions.

\section{Interaction potentials, the Hamiltonian and the rates}

In this section we develop the framework that will be used as the starting point for the
analysis of the NSI couplings in section 3. To this end we review the derivation of the 
neutrino interaction potentials in solar matter, its generalization to non standard 
interactions along with the corresponding matter Hamiltonian and the event rates.

\subsection{Interaction potentials and the Hamiltonian}

While $\nu_e$'s propagate through solar matter their interaction with electrons proceeds
both through charged and neutral currents (CC) and (NC). Recalling that for standard 
interactions (SI) each tree level vertex accounts for a factor
\begin{equation}
\frac{g_L}{cos\theta_W}(T_{3L}-2Q_f sin^2\theta_{W}),
\end{equation}
inserting the W, Z propagators and the electron external lines, one gets for the $\nu_e$ 
interaction potentials
\begin{equation}
(V_e)_{CC}=G_F\sqrt{2}N_e~,~(V_e)_{NC}=G_F/\sqrt{2}(-1+4sin^2\theta_W)N_e
\end{equation}
where $G_F$ is the Fermi constant, $G_F/\sqrt{2}=g_L^2/8m^2_W$. For the interactions
with quarks only neutral currents are involved and the additivity of the quark-current 
vertices gives for protons
\begin{equation}
V_p=(V_p)_{NC}=G_F/\sqrt{2}(1-4 sin^2 \theta_W)N_e
\end{equation}
and for neutrons
\be
V_n=(V_n)_{NC}=-G_F/\sqrt{2}N_n.
\ee
Hence the neutrino interaction potential is for standard interactions \footnote{All expressions
are divided by 2 to account for the fact that the medium is unpolarized.}
\be
V(SI)=V_e+V_p+V_n=G\sqrt{2}N_e\left(1-\frac{N_n}{2N_e}\right)=
V_c+V_n
\ee
with $V_e=(V_e)_{CC}+(V_e)_{NC}$ and $V_c=V_e+V_p=G_F\sqrt{2}N_e$.

In order to introduce NSI we assume that each diagram associated to neutrino
propagation in matter (i.e. CC and NC currents in $\nu_{\alpha}~e^{-}$ and NC
currents in $\nu_{\alpha}~u,~\nu_{\alpha}~d$ scattering) is multiplied by a factor
$\varepsilon_{\alpha\beta}^{e,u,d~P}$ parameterising the deviation from the standard model.
So we assume that the interaction potential for $\nu_{\alpha}$ ($\alpha=e,~\mu,~\tau)$ on 
electrons involves both CC and NC giving rise to possible lepton flavour violation: 
$\nu_{\alpha}$ for $\alpha \neq e$ {\it may} have CC. So for the charged current of $\nu_e$
with electrons we have
\be
(V_{e})_{CC}(NSI)=\frac{g^2_L}{2m^2_W}(\varepsilon_{\alpha \beta}^{eP})_{CC}N_e
\ee
and for the neutral current
\be
(V_{e})_{NC}(NSI)=\frac{g^2_L}{4m^2_W}(-1+4sin^2\theta_W)(\varepsilon_{\alpha \beta}^{eP})_{NC}N_e
\ee
where the NSI couplings affecting the CC and NC processes should in principle be distinguished.

Using equation (2) and the additivity of the quark-current vertices, one gets for the 
neutrino interaction potential with protons 
\be
V_p(NSI)=\frac{g^2_L}{2m^2_W}\left[\varepsilon_{\alpha \beta}^{uP}-
\frac{\varepsilon_{\alpha \beta}^{dP}}{2}-\left(\frac{4}{3}2\varepsilon_{\alpha \beta}^{uP}-
\frac{2}{3}2\varepsilon_{\alpha \beta}^{dP}\right)sin^2\theta_W\right]N_e.
\ee
Similarly for neutrons
\be
V_n(NSI)=\frac{g^2_L}{2m^2_W}\left(\frac{\varepsilon_{\alpha \beta}^{uP}}{2}-
\varepsilon_{\alpha \beta}^{dP}\right)N_n.
\ee
In both (9) and (10) only neutral currents are involved.

Adding (7), (8), (9) and (10) and dividing by 2 one finally gets
\begin{eqnarray}
V(NSI) & = & G\sqrt{2}N_e\left[(\varepsilon_{\alpha \beta}^{eP})_{CC}+
\left(-\frac{1}{2}+2sin^2\theta_W \right)
(\varepsilon_{\alpha \beta}^{eP})_{NC}+\left(1-\frac{8}{3}sin^2\theta_W+\frac{N_n}{2N_e} \right) 
\varepsilon_{\alpha \beta}^{uP}\right.\nonumber\\
&& + \left.\left(-\frac{1}{2}+\frac{2}{3}sin^2\theta_W-\frac{N_n}{N_e} \right)
\varepsilon_{\alpha \beta}^{dP} \right]
\end{eqnarray}
which reduces to eq.(6) in the absence of NSI. 

In the case of the standard interactions, the interaction potentials for $\nu_e$ and 
$\nu_{\alpha}$ constitute a diagonal matrix because they cannot be responsible for flavour 
change [eq.(6)]. This may occur as a consequence of the vacuum mixing angle
(oscillations) \cite{Fogli:2008ig,Schwetz:2008er} or the magnetic moment for 
instance \cite{Das:2009kw,Pulido:2009sb}. On the other hand, in the case of NSI 
the interaction potentials [eq.(11)] constitute a full matrix in neutrino flavour space. 

In order to obtain the matter Hamiltonian eqs.(6) and (11) must now be added. In the 
flavour basis this is
\be
{\cal H}_M=V_c \left(\begin{array}{ccc} 1 & 0 & 0 \\
0 & 0 & 0 \\
0 & 0 & 0 \\ \end{array}\right)+
\left(\begin{array}{ccc} v_{ee}(NSI) & v_{e\mu}(NSI) & v_{e\tau}(NSI) \\
v_{\mu e}(NSI) & v_{\mu\mu}(NSI) & v_{\mu\tau}(NSI) \\
v_{\tau e}(NSI) & v_{\tau\mu}(NSI) & v_{\tau\tau}(NSI) \\ \end{array}\right)~
\ee
where in the first term, describing the standard interactions, the additive quantity $V_n$
which is proportional to the identity, has been removed from the diagonal. In the second term
$v_{\alpha\beta}$ ($\alpha,\beta=e,\mu,\tau$) denote the matrix elements of the interaction 
potential matrix (11). Finally in the mass basis
\be
{\cal H}=\left(\begin{array}{ccc} 0 & 0 & 0 \\
0 & \frac{\Delta m^2_{21}}{2E} &  0 \\
0 &       0         &  \frac{\Delta m^2_{31}}{2E} \\ \end{array} \right)+ U^{\dagger}
{\cal H}_M U
\ee
where $U$ is the PMNS matrix \cite{Maki:1962mu} \footnote{We use the standard 
parameterization \cite{Amsler:2008zzb} for the $U$ matrix and the central value 
$sin\theta_{13}=0.13$ claimed in ref. \cite{Fogli:2008ig}.}, $E$ is the neutrino energy
and $\Delta m^2_{ij}=m_i^2-m_j^2$ with $m_i$ ($i=1,2,3$) the neutrino mass. Upon
insertion of this Hamiltonian expression in the neutrino evolution equation, the
survival ($P_{ee}$) and conversion probabilities ($P_{e\mu},P_{e\tau}$) are evaluated 
using the Runge-Kutta numerical integration.

\subsection{Neutrino electron scattering detection rates}

For the detection in SuperKamiokande and SNO through $\nu_{\alpha}~e^{-} \rightarrow 
\nu_{\beta}~e^{-}$ scattering, the NSI information comes in the probabilities and the cross 
section 
\be
\frac{d\sigma}{dT}=\frac{2G_F^2 m_e}{\pi}\left[\tilde g_L^2+\tilde g_R^2\left(1-
\frac{T}{E_{\nu}}\right)^2-\tilde g_L \tilde g_R\frac{m_e T}{E^2_{\nu}}\right]
\ee
where $\tilde g_{L,R}$ are the $g_{L,R}$ couplings modified according to \cite{Barranco:2005ps}
\begin{eqnarray*} 
(\tilde g_{L,R})_{\nu_e}^2&=&\left|(g_{L,R})_{\nu_e}+\varepsilon_{ee}^{L,R}\right|^2+
|\varepsilon_{\mu e}^{L,R}|^2+
|\varepsilon_{\tau e}^{L,R}|^2~~~~~~{\rm for}~~~~~~\nu_e~e^{-}
\rightarrow \nu_{\alpha}~e^{-} \\
(\tilde g_{L,R})_{\nu_{\mu}}^2&=&\left|(g_{L,R})_{\nu_{\mu}}+\varepsilon_{\mu\mu}^{L,R}\right|^2+
|\varepsilon_{e\mu}^{L,R}|^2+
|\varepsilon_{\tau\mu}^{L,R}|^2~~~~~~{\rm for}~~~~~~\nu_{\mu}~e^{-}
\rightarrow \nu_{\alpha}~e^{-} \\
(\tilde g_{L,R})_{\nu_{\tau}}^2&=&\left|(g_{L,R})_{\nu_{\tau}}+\varepsilon_{\tau\tau}^{L,R}\right|^2+
|\varepsilon_{e\tau}^{L,R}|^2+
|\varepsilon_{\mu\tau}^{L,R}|^2~~~~~~{\rm for}~~~~~~\nu_{\tau}~e^{-}
\rightarrow \nu_{\alpha}~e^{-}~.
\end{eqnarray*}
with $\alpha=e,\mu,\tau$.

For $\nu_e$ both charged and neutral currents are possible, so that
\be
(g_L)_{\nu_e}=\frac{1}{2}+sin^2\theta_W~,~~(g_R)_{\nu_e}=sin^2\theta_W
\ee
whereas $\nu_{\mu,\tau}$ only interact through neutral currents, hence
\be
(g_L)_{\nu_{\mu},\nu_{\tau}}=-\frac{1}{2}+sin^2\theta_W~,~~(g_R)_{\nu_{\mu},\nu_{\tau}}=
sin^2\theta_W.
\ee


These expressions are then inserted in the spectral event rate 
\begin{equation}
R^{th}_{SK,SNO}(E_e)\!\!=\!\!
\frac{\displaystyle\int_{m_e}^{{E'_e}_{max}}\!\!dE'\!\!_ef(E'_e,E_e)
\!\!\int_{E_m}^{E_M}\!\!dE\phi(E)\!\!\left[P_{ee}(E)\frac{d\sigma_{e}}{dT'}\!+\!
P_{e\mu}(E)\frac{d\sigma_{\mu}}{dT'}\!+\!P_{e\tau}(E)\frac{d\sigma_{\tau}}{dT'}\right]}
{\displaystyle\int_{m_e}^{{E'_e}_{max}}\!\!dE'\!\!_e
f(E'_e,E_e)\!\!\int_{E_m}^{E_M}\!\!dE\phi(E)\frac{d\sigma_{e}}{dT'}}
\end{equation}
which will be evaluated in the next section. Here $\phi(E)$ denotes the neutrino flux 
from Boron and hep neutrinos, $f(E^{'}_e,E_e)$ is the energy resolution function for
SuperKamiokande and SNO \cite{Fukuda:1998fd,Aharmim:2005gt} and the rest of the notation 
is standard. 

Notice that whereas the Hamiltonian (13) is symmetric under the interchange 
$$\varepsilon_{\alpha\beta}^L \leftrightarrow \varepsilon_{\alpha\beta}^R~~~{\rm for}~~~e,u,d$$
such is not the case for the detection process [see eq.(14)-(16)]. We finally note that at
the detection level the NSI couplings $\varepsilon_{\alpha\beta}^{L,R}$ are considered 
separately, as clearly seen from eqs.(14), whereas at the level of propagation, since the 
diagrams involved in the interaction potentials add up, their sum should instead be 
considered.

\section{NSI couplings, probabilities and spectra}

We now perform an investigation of the effect of the NSI couplings $\varepsilon_
{\alpha\beta}^{e,u,d}=|\varepsilon_{\alpha\beta}^{e,u,d}|e^{i\phi_{\alpha\beta}^{e,u,d}}$ 
on the neutrino probability and event rates. Our aim is to find those couplings which
lead to a flat SuperKamiokande spectral rate, thus improving the fit with respect to its 
LMA prediction while keeping the quality of the other solar event rate fits. We first 
consider equal CC and NC couplings, namely $(\varepsilon_{\alpha \beta}^{eP})_{CC}=
(\varepsilon_{\alpha \beta}^{eP})_{NC}=\varepsilon_{\alpha \beta}^{eP}$ and in a second
stage $(\varepsilon_{\alpha \beta}^{eP})_{CC}\neq (\varepsilon_{\alpha \beta}^{eP})_{NC}$.

\subsection {$(\varepsilon_{\alpha \beta}^{eP})_{CC}=(\varepsilon_{\alpha \beta}^{eP})_{NC}
=\varepsilon_{\alpha \beta}^{eP}$}

For the sake of clarity we will organize the NSI couplings in three matrices according to 
whether the charged fermion in the external line is $e,~u,~d$
\be
\left(\begin{array}{ccc} \varepsilon_{ee}^{e,u,d~P} & \varepsilon_{e\mu}^{e,u,d~P} & 
\varepsilon_{e\tau}^{e,u,d~P} \\
\varepsilon^{*e,u,d~P}_{e\mu} & \varepsilon_{\mu\mu}^{e,u,d~P} & 
\varepsilon_{\mu\tau}^{e,u,d~P} \\
\varepsilon^{*e,u,d~P}_{e\tau} & \varepsilon^{*e,u,d~P}_{\mu\tau} & 
\varepsilon_{\tau\tau}^{e,u,d~P} \\ \end{array}\right).
\ee
Each set of three couplings $\varepsilon_{\alpha\beta}^{e,u,d~P}$ enters in equation
(11) in the entry $v_{\alpha\beta}$ of the interaction potential matrix. Altogether 
there are 18 couplings with 36 parameters: each matrix 
of the three in eq.(18) contains 6 independent entries, each with a modulus and a phase. 
We analyse one coupling at a time, by taking all others zero. We first consider the 
cases of purely real and imaginary couplings, hence 4 possibilities for each phase 
\be 
\phi_{\alpha\beta}^{e,u,d}=0,\pi/2,\pi,(3/2)\pi.
\ee
Motivated by the arguments expound in the introduction we investigate the parameter range
$|\varepsilon_{\alpha\beta}|~\epsilon~[5\times 10^{-5}~,~5\times 10^{-2}]$. We find that 

\begin{itemize}
\item Off diagonal entries $\varepsilon_{\alpha\beta}^{e,u,d~P}$ ($\alpha\neq\beta$) which
contain 3$\times$3$\times$4$=$36 possibilities for moduli and phases do not induce any
change in the LMA probability, nor any visible change in the rates, either if one or more
at a time are inserted.

\item Diagonal entries $\varepsilon_{\alpha\alpha}^{e,u,d~P}$. 

(a) Real couplings $\varepsilon_{\alpha\alpha}^{e,u,d~P}=\pm|\varepsilon_{\alpha\alpha}^{e,u,d~P}|$
(3$\times$3$\times$2$=$18 possibilities) do not change the LMA probability, hence the rates.

(b) Imaginary couplings $\varepsilon_{\alpha\alpha}^{e,u,d~P}=
\pm i |\varepsilon_{\alpha\alpha}^{e,u,d~P}|$ (3$\times$3$\times$2$=$18 possibilities)  
lead to probabilities which diverge from $P_{LMA}$ for all $|\varepsilon_{\alpha\alpha}|>
5\times 10^{-5}$. According to the probability shape that is obtained, we group these cases in 
the following way
\end{itemize}

\vspace{-0.6cm}

\begin{center}
\begin{tabular}{c|ccc}  
  &         1                &           2               &           3         \\ \hline \hline
A & $+i|\varepsilon_{ee}^{e~P}|$ & $+i|\varepsilon_{\mu\mu}^{e~P}|$ &  $-i|\varepsilon_{ee}^{e~P}|$ \\
B & $+i|\varepsilon_{ee}^{u~P}|$ & $+i|\varepsilon_{\mu\mu}^{u~P}|$ &  $-i|\varepsilon_{ee}^{u~P}|$ \\
C & $-i|\varepsilon_{ee}^{d~P}|$ & $-i|\varepsilon_{\mu\mu}^{d~P}|$ &  $+i|\varepsilon_{ee}^{d~P}|$ \\ \hline
D & $-i|\varepsilon_{\mu\mu}^{e~P}|$ & $-i|\varepsilon_{\tau\tau}^{e~P}|$ &  $+i|\varepsilon_{\tau\tau}^{e~P}|$ \\ 
E & $-i|\varepsilon_{\mu\mu}^{u~P}|$ & $-i|\varepsilon_{\tau\tau}^{u~P}|$ &  $+i|\varepsilon_{\tau\tau}^{u~P}|$ \\ 
F & $+i|\varepsilon_{\mu\mu}^{d~P}|$ & $+i|\varepsilon_{\tau\tau}^{d~P}|$ &  $-i|\varepsilon_{\tau\tau}^{d~P}|$ \\ 
\end{tabular}
\end{center}
\begin{center}
{\it{Table I - The NSI couplings that modify the LMA probability
.}}
\end{center}

\noindent All cases in the first column of table I along with D2, E2, F2 in 
the second column lead qualitatively to the same monotonically decreasing probability curve: 
a high probability ($P\geq P_{LMA}$) for low energy ($E\lesssim 3 MeV$) and a low one 
($P\leq P_{LMA}$) for intermediate and high energies. The curve becomes  
increasingly flat in this energy sector as $\varepsilon_{\alpha\alpha}$ increases, 
which is also reflected in the flatness of the SuperKamiokande spectral rate.
However for cases A1, B1, C1 and D2, E2, F2 the probability gets too high for low 
energies so that the Ga \cite{Cattadori:2005gq,Gavrin:2007wc} rate fails to be 
conveniently fitted. The 'best' results in the sense that they 
lead to the most flat spectral rate which approaches the SuperKamiokande one and 
to a correct fit for Ga are obtained alternatively from cases D1, E1, or F1 for the 
following values

\be
\begin{array}{ccc}
-i|\varepsilon_{\mu\mu}^{e~P}| & = & -i~1.5\times 10^{-3}  \\
-i|\varepsilon_{\mu\mu}^{u~P}| & = & -i~2.5\times 10^{-3}  \\
+i|\varepsilon_{\mu\mu}^{d~P}| & = & +i~2.0\times 10^{-3}  .
\end{array}
\ee
For larger values of the NSI couplings the probability moves further away from its LMA profile 
so that the Ga rate becomes too high and the $^8 B$ one too low. In fig.1 we plot four 
survival probabilities: at the lowest energy and from bottom to top the first curve is the
LMA one, the next corresponds to all cases in (20) and leads to the best fit of the four, the 
next one to the case 
$-i|\varepsilon_{\mu\mu}^{e~P}|=-i~3\times 10^{-3}$ and the top one to $+i|\varepsilon_{ee}^{e~P}|=
+i~5\times 10^{-3}$. A comparison is shown in table II between the predictions and the quality
of the fits obtained from LMA and the case $-i|\varepsilon_{\mu\mu}^{e~P}|=-i~1.5\times 10^{-3}$.
In figs.2 and 3 we show the SuperKamiokande spectrum for LMA (upper curves) and for the first case 
in (20) (lower curves) superimposed on the data points taken respectively from refs.
\cite{Fukuda:2002pe} and \cite{:2008zn}. The improvement obtained through the NSI coupling 
is clearly visible. In fig.4 the two curves are superimposed on 
the SNO data points for electron scattering \cite{Aharmim:2009gd}. Here the data are also 
clearly consistent with a constant rate.

\vspace{-0.6cm}

\begin{center}
\begin{tabular}{ccccccccccc} \\ \hline \hline
     & Ga & Cl & SK & $\rm{SNO_{NC}}$ & $\rm{SNO_{CC}}$ & $\rm{SNO_{ES}}$ &
$\!\!\chi^2_{rates}\!\!$ & $\chi^2_{{SK}_{sp}}$ & $\chi^2_{SNO}$ & $\chi^2_{gl}$\\ \hline
LMA  & 64.9 & 2.84 & 2.40 & 5.47 & 1.79 & 2.37 & 0.67 & 42.0 & 48.6 & 91.3 \\
$-i|\varepsilon_{\mu\mu}^{e~P}|$ & 69.7 & 2.74 & 2.23 & 5.47 & 
1.68 & 2.26 & 0.11 & 40.3 & 45.0 & 85.4 \\ \hline
\end{tabular}
\end{center}
{\it{Table II - Comparison between the LMA predictions for solar event rates and the NSI ones 
with $-i|\varepsilon_{\mu\mu}^{e~P}|=-i~1.5\times 10^{-3}$. For details of the $\chi^2$ analysis
see for instance \cite{Das:2009kw}.}}

\vspace{0.5cm}

The first set of cases in the second column of table I, namely A2, B2, C2, lead qualitatively 
to the inverse behaviour with energy of the LMA probability. As $|\varepsilon_{\alpha\alpha}|$ 
increases from its lower bound, one gets $P\leq P_{LMA}$ for low energies ($E\lesssim 2-3 MeV$) 
and $P\geq P_{LMA}$ for intermediate and high energies, so that the fits worsen with respect to 
the LMA ones.

Finally all cases in the third column of table I, namely A3, B3, C3, D3, E3, F3 lead to
probability curves which are totally unsuitable: they deviate drastically from both $P_{LMA}$ 
and a from flat, suitable profile able to generate the SuperKamiokande spectrum.

We have also checked that combinations of real and imaginary parts for all couplings do 
not change the previous results. This should be expected since, as mentioned earlier, 
purely real couplings do not change the LMA probability. The only consequence of introducing
real parts in the NSI couplings comes in the spectral event rates through the quantities
$\tilde g_{L,R}$ in eqs.(14),(15), but the differences lie much beyond the experimental
accuracy.

\subsection {$(\varepsilon_{\alpha \beta}^{eP})_{CC}\neq
(\varepsilon_{\alpha \beta}^{eP})_{NC}$}

The analysis of the more general case of different CC and NC couplings affecting the 
$\nu_e~e$ scattering diagrams can be made quite simple if one examines the coefficients of 
$(\varepsilon_{\alpha \beta}^{eP})_{CC}$ and $(\varepsilon_{\alpha \beta}^{eP})_{NC}$ in
eq.(11). The first is unity whereas for equal CC and NC couplings it is 0.96 and the 
second is now -0.04 as compared to the previous value 0.96 as well. Consequently one 
expects that the analysis for $(\varepsilon_{\alpha \beta}^{eP})_{CC}$ leads to 
approximately the same results as for equal couplings while the results are modified by
a factor of 0.96/(-0.04) in the analysis for $(\varepsilon_{\alpha \beta}^{eP})_{NC}$.
Indeed the convenient modification in the LMA probability is obtained for 
\be
-i(|\varepsilon_{\mu\mu}^{e~P}|)_{CC} = -i~1.4\times 10^{-3}
\ee
or alternatively
\be
+i(|\varepsilon^{e~P}_{\mu\mu}|)_{NC}=+i~3.6\times 10^{-2}.
\ee
which lead to the same probability as the cases listed in eq.(20). The results for the 
other couplings involving u and d quarks are of course unchanged. As before we have 
considered one coupling at a time to be non zero. 

\begin{figure} [htb]
\centering
\includegraphics[height=160mm,keepaspectratio=true,angle=270]{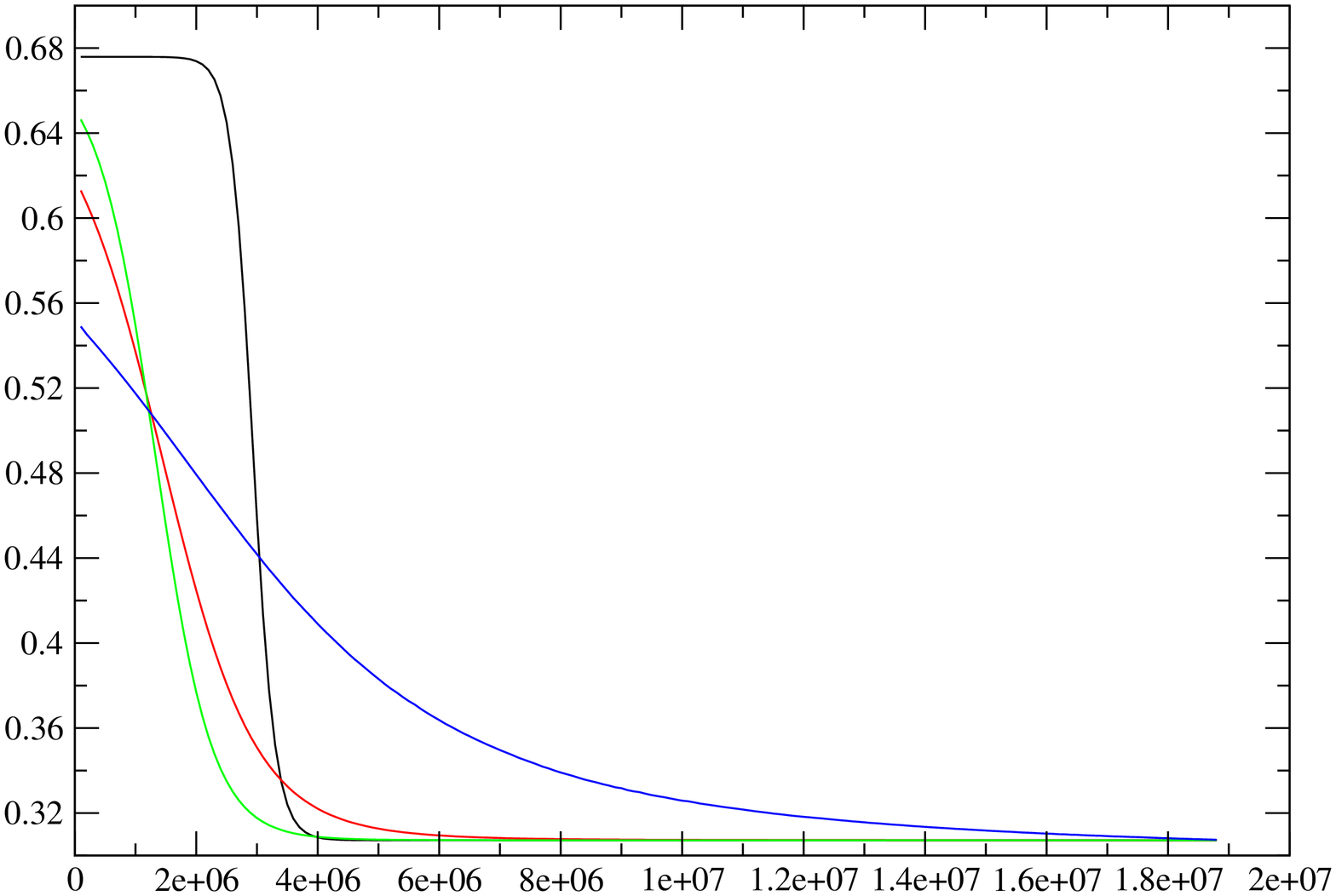}
\caption{ \it Survival probabilities as a function of neutrino energy in eV. At the lowest
energy and from bottom to top, the first curve is the LMA one, the next is the one providing
the best fit to the data with $-i|\varepsilon_{\mu\mu}^{e~P}|=-i~1.5\times 10^{-3}$ [eq.(20)], 
the next is for $-i|\varepsilon_{\mu\mu}^{e~P}|=-i~3\times 10^{-3}$ and the top one is for
$-i|\varepsilon_{\mu\mu}^{e~P}|=+i~5\times 10^{-3}$ with other non standard couplings vanishing 
in each case. The last two curves lead to an unacceptably high Ga rate prediction 
\cite{Cattadori:2005gq,Gavrin:2007wc}.}
\end{figure}

\section{Conclusions}

We have investigated the prospects for improving the LMA predictions for solar neutrino
event rates with NSI. At present there is no evidence of any new physics associated to a 
scale not too far above the electroweak scale, hence the great variety of theoretical models 
available for NSI. In our approach we assumed that NSI are extra contributions 
to the vertices $\nu_{\alpha}\nu_{\beta}$ and $\nu_{\alpha}e$, so the new couplings describe 
the deviation from the standard model. With this in mind we derived the neutrino interaction 
potential in solar matter which was added to the standard Hamiltonian, proceeding with the
integration of the evolution equation through the Runge-Kutta method. Neutral and charged
current couplings are involved in interactions with electrons whereas only neutral couplings
affect those with quarks. We considered the new interactions both at the propagation and at 
the detection level. The improvement we searched for the LMA predictions consisted in 
finding whether and how the modification induced by NSI can lead to a flat spectral event 
rate for SuperKamiokande and an event rate for the Cl experiment within 1$\sigma$ of the 
data, while keeping the accuracy of all other predictions.

We used the current notation for the NSI couplings $\varepsilon_{\alpha\beta}^{e,u,d~P}$ where
$\alpha,\beta$ are the neutrino labels and $e,u,d$ denote the charged fermion involved in the
process. We investigated the range $|\varepsilon_{\alpha\beta}|~\epsilon~[5\times 
10^{-5}~,~5\times 10^{-2}]$ and may summarize our main results as follows

\begin{itemize}
\item Real couplings $\varepsilon_{\alpha\beta}^{e,u,d~P}=\pm|\varepsilon_{\alpha\beta}
^{e,u,d~P}|$ do not change the LMA probability when considered either one at a time or 
altogether and thus they induce a small change in the neutrino electron scattering rate 
($\lesssim 1\%$) which is far beyond experimental visibility.

\item Off diagonal couplings $\varepsilon_{\alpha\beta}^{e,u,d~P}$ ($\alpha\neq \beta$) 
considered either one at a time or altogether do not change the LMA probability and thus the rates 
in a significant way.

\item Diagonal, imaginary couplings $\varepsilon_{\alpha\alpha}^{e,u,d~P}= 
\pm i |\varepsilon_{\alpha\alpha}^{e,u,d~P}|$ are the only ones that lead to changes of all 
kinds in the probability and hence the rates.

\item The couplings that lead to the 'best' results in the sense of leading to a flat
spectral SuperKamiokande rate and a good fit for all other rates are as
follows
\be
\begin{array}{ccc}
-i|\varepsilon_{\mu\mu}^{e~P}| & = & -i~1.5\times 10^{-3}  \\
-i|\varepsilon_{\mu\mu}^{u~P}| & = & -i~2.5\times 10^{-3}  \\
+i|\varepsilon_{\mu\mu}^{d~P}| & = & +i~2.0\times 10^{-3}  
\end{array}
\ee
within the assumption of equal charged (CC) and neutral current (NC) couplings.

\item If one assumes different (CC) and (NC) couplings we have
\be
\begin{array}{ccc}
-i(|\varepsilon_{\mu\mu}^{e~P}|)_{CC} & = & -i~1.4\times 10^{-3} \\
+i(|\varepsilon^{e~P}_{\mu\mu}|)_{NC} & = & +i~3.6\times 10^{-2}.
\end{array}
\ee
\end{itemize}
In both cases (23) and (24) the results should be taken alternatively.

Since real couplings do not change the LMA probability, the insertion of real parts 
in the couplings does not lead to any change in the above results.
We also note that whereas an increase in $sin\theta_{13}$ leads to a mere decrease in the 
$^8 B$ spectral rate which translates in a parallel shift of the curve \cite{Das:2009kw},
a convenient choice of NSI couplings [eqs.(23) or (24)] induces in contrast an increased 
flatness. The situation is similar to the magnetic moment conversion to sterile neutrinos
\cite{Das:2009kw}.

Unless stated otherwise our analyses were done for one coupling at a time. There are of
course infinite combinations of couplings so that a complete investigation is not possible 
and would unlikely lead to new conclusions. The aim of this paper is to prove that
NSI provide a true possibility to solve the remaining inconsistencies within the solar
neutrino problem.

\vspace{1cm}
\noindent {\Large \bf Acknowledgements}
\vspace{0.5cm}

{\em We are grateful to Marco Picariello for useful discussions.
C. R. Das gratefully acknowledges a scolarship from Funda\c{c}\~{a}o para
a Ci\^{e}ncia e Tecnologia ref. SFRH/BPD/41091/2007.} 


\begin{figure} [htb]
\centering
\includegraphics[height=150mm,keepaspectratio=true,angle=270]{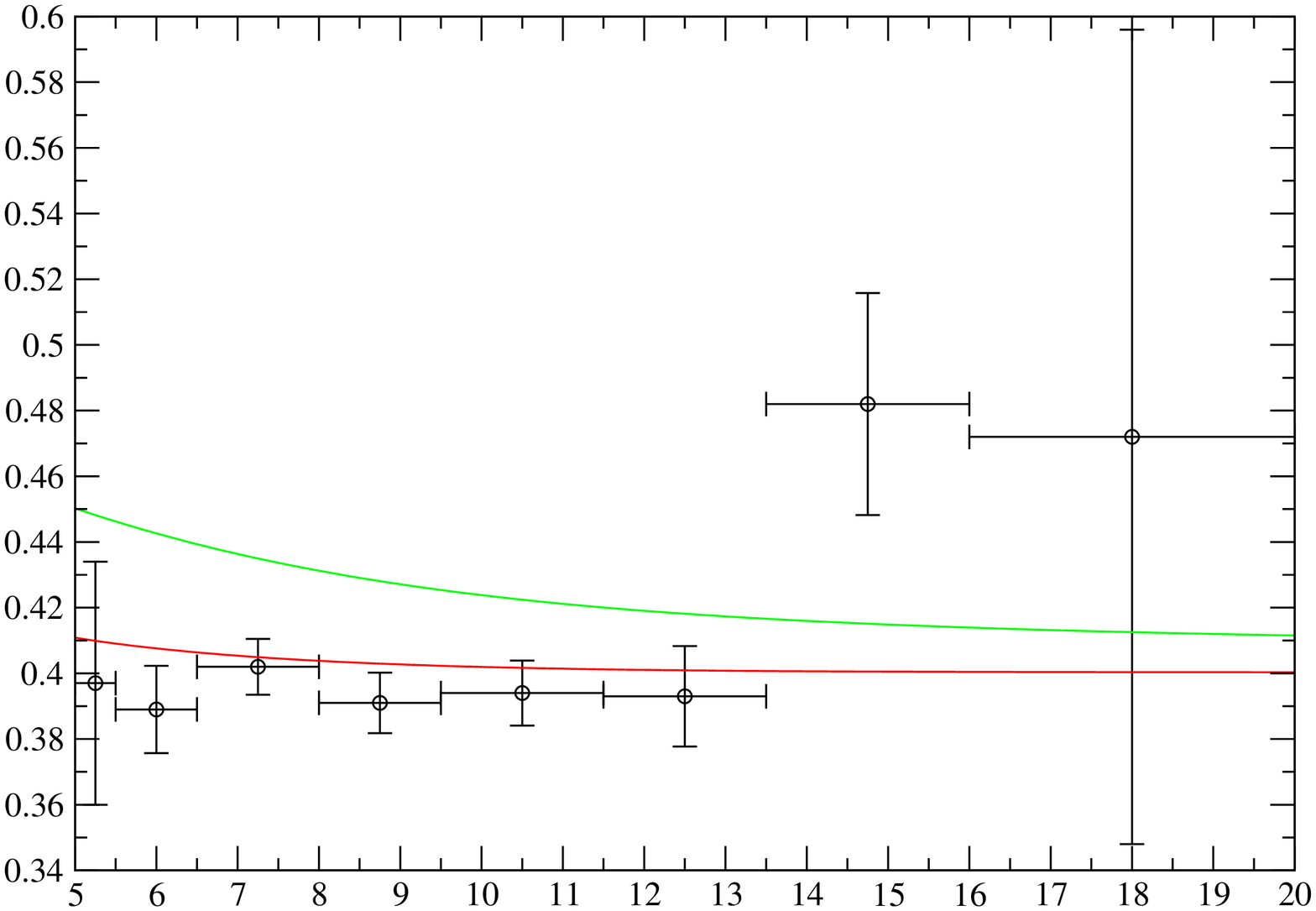}
\caption{ \it Predictions for SuperKamiokande (units in MeV for electron energy). 
The upper curve is the LMA spectrum and the lower curve is the LMA spectrum with non 
standard interactions as in eqs.(23) or (24). These are superimposed on the data 
published by the Collaboration in 2002 \cite{Fukuda:2002pe}.}
\end{figure}

\begin{figure} [htb]
\centering
\includegraphics[height=150mm,keepaspectratio=true,angle=270]{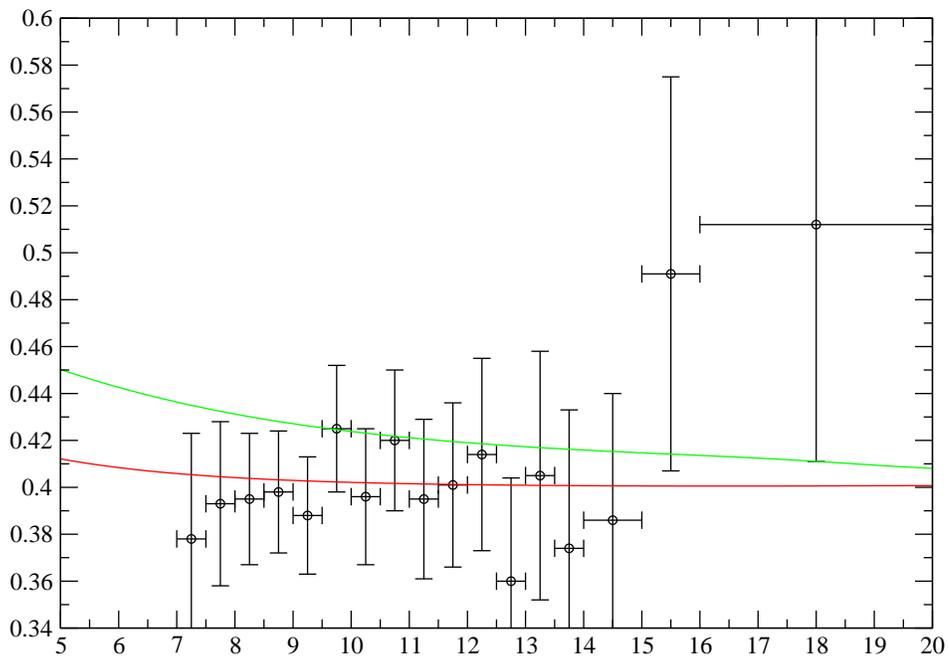}
\caption{\it Same as fig.2 with the data published in 2008 \cite{:2008zn}.}
\end{figure}

\begin{figure} [htb]
\centering
\includegraphics[height=150mm,keepaspectratio=true,angle=270]{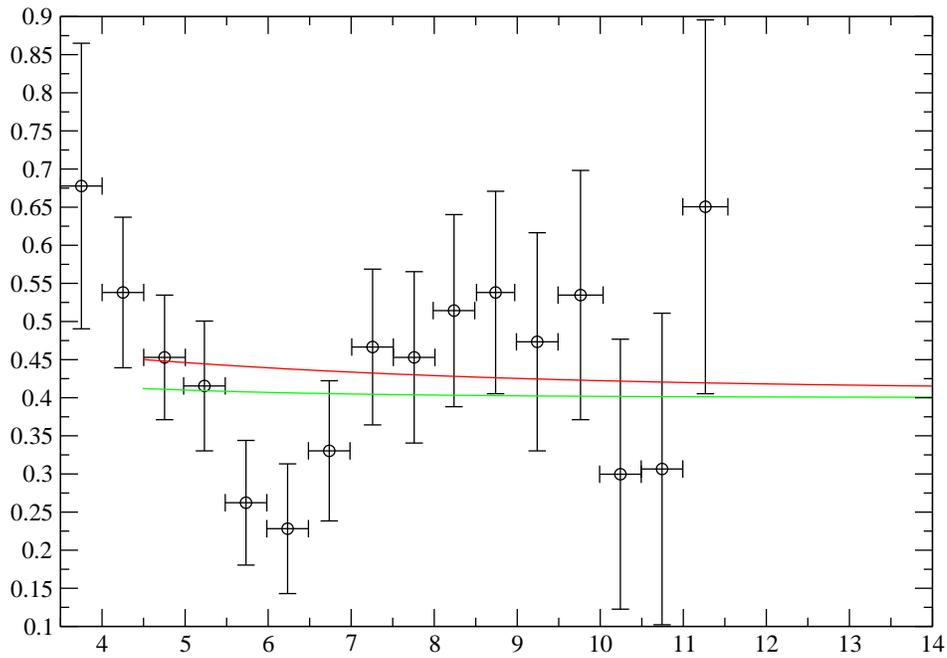}
\caption{\it Predictions for SNO neutrino electron scattering  
superimposed on the data \cite{Aharmim:2009gd} (units in MeV for electron kinetic 
energy). The upper curve is the LMA prediction and the lower curve is the LMA one 
with non standard interactions. Errors bars are larger than in SuperKamiokande so 
that the data are consistent with a flat spectrum.}
\end{figure}

\end{document}